\documentclass{PoS}

\usepackage{graphicx}
\def\CP{$ C \! P$}

\def\ra{\rightarrow}
\def\mes{$m_{ES}$}
\def\DE{$\Delta E$}
\def\fisher{${\cal F}$}
\def\Bkpp{$B^+ \ra K^+ \pi^+ \pi^- \gamma$}
\def\Bkspp{$B^0 \ra K^0_S \pi^+ \pi^- \gamma$}
\def\mkpp{$m_{K \pi \pi}$}
\def\mkp{$m_{K\pi}$}
\def\mpp{$m_{\pi \pi}$}
\def\splot{$_s{\cal P}lot$}
\def\Ktwelve{$K_1(1270)$}
\def\Kfourteenz{$K_1(1400)$}
\def\Kfourteent{$K^*(1410)$}
\def\Ksixteen{$K^*(1680)$}
\def\Kfourteenth{$K^*_2(1430)$}

\def\babar{\mbox{{\normalsize \sl B}\hspace{-0.4em} {\small \sl A}\hspace{-0.03em}{\normalsize \sl B}\hspace{-0.4em} {\small \sl A\hspace{-0.02em}R}}}

\newcommand\T{\rule{0pt}{2.6ex}}       
\newcommand\B{\rule[-1.2ex]{0pt}{0pt}} 

\title{Study of $B \ra K \pi \pi \gamma $ Decays}

\ShortTitle{$B \ra K \pi \pi \gamma $ Decays}

\author{\speaker{G. Eigen} \\ representing the \babar\ collaboration\thanks{This work is supported by the Norwegian Research Council.}\\
        Dept. of Physics, University of Bergen, Bergen, Norway \\
        E-mail: \email{gerald.eigen@ift.uib.no}}

\abstract{Using $471 \times 10^6 ~B \bar B$ decays recorded with the \babar\ detector at the PEP-II $e^+ e^-$ storage ring, we present the time-dependent \CP\ asymmetry measurement in the radiative penguin decay mode $B^0 \ra K^0_S \rho(770)^0 \gamma$, yielding $S_{K^0_S \rho^0 \gamma} =-0.17\pm 0.32^{+0.07}_{-0.06}$. The result is extracted from the time-dependent \CP~asymmetry parameters  $S_{K^0_S \pi^+ \pi^- \gamma} =0.14\pm 0.25^{+0.04}_{-0.03}$ and $C _{K^0_S \pi^+ \pi^- \gamma} =-0.39\pm \pm 0.20\pm 0.05$ measured in the neutral decay $B^0 \ra K^0_S \pi^+ \pi^- \gamma$ after correcting for the dilution of $K^*(892) \pi \gamma$ in $K \rho \gamma$. The dilution factor  $ D_{K^0_S \rho \gamma} =-0.79^{+0.18}_{-0.17}$ is determined from a study of the charged mode $B^+ \ra K^+ \pi^+\pi^-\gamma$, which produces more signal events and is related to the neutral mode by isospin. We need a detailed knowledge of the resonance structure in the $K^+ \pi^+ \pi^-$ mass spectrum and measure branching fractions of different resonances to $K \pi$ and $\pi \pi$ final states. 
We also measure the branching fractions ${\cal B} (B^+ \ra K^+ \pi^+ \pi^- \gamma)= (27.2\pm 1.0 \pm 1.2) \times 10^{-6}$ and ${\cal B} (B^0 \ra K^0_S \pi^+ \pi^- \gamma)= (24.0\pm 2.4^{+1.7}_{-1.8}) \times 10^{-6}$. }

\FullConference{The European Physical Society Conference on High Energy Physics\\
		22--29 July 2015\\
		Vienna, Austria}

\begin{document}

\section{Introduction}

The V-A structure of the Standard Model (SM) weak interaction produces predominantly left-handed photons in $b \ra s \gamma $ decays.
Thus apart from $m_s /m_b$ effects~\cite{Atwood}, a $B~(\bar B)$ meson decays predominantly to a right-handed (left-handed) photon. In the SM, the mixing-induced \CP\ asymmetry in $B \ra f_{CP} \gamma$ decays is expected to be small in the SM where $f_{CP}$ is a \CP\ eigenstate. However, new physics processes in which opposite-helicity photons are involved may alter the SM prediction~\cite{Atwood, Fujikawa, Babu, Cho}. 
Inclusive and exclusive radiative decays have been studied by \babar~\cite{babar12, babar01} and Belle~\cite{belle09, belle04} in several channels. For example, the inclusive branching fraction of ${\cal B}(B \ra X_s \gamma)=(3.40\pm 0.21)\times 10^{-4} $~\cite{pdg} is in good agreement with the SM prediction of ${\cal B}(B \ra X_s \gamma)=(3.15\pm 0.23)\times 10^{-4} $~\cite{misiak}. 
\babar\ has studied the exclusive decay $B \ra K \pi \pi \gamma $ exploring the resonance structure of the $K \pi \pi$ system.
The data sample consists of $471 \times 10^6 ~B \bar B$ events recorded with the \babar\ detector at the PEP II $e^+ e^-$ storage ring at SLAC corresponding to an integrated luminosity of 426 $\rm fb^{-1}$. 

The goal consists of measuring the mixing-induced \CP\ asymmetry parameter, $S_{K^0 \rho^0 \gamma}$, in the $B$ radiative decay to the \CP\ eigenstate $K^0_S \rho^0 \gamma$, which is sensitive to right-handed photons. To accomplish this we measure the time-dependent \CP\ asymmetry parameters $S_{K^0 \pi \pi \gamma}$ and $C_{K^0 \pi \pi \gamma}$ in the neutral decay $B^0 \ra K^0_S \pi^+ \pi^- \gamma$ since $S_{K^0_S \rho \gamma} =\frac{S_{K^0_S \pi^+ \pi^- \gamma} }{D_{K^0_S \rho \gamma} } $. The dilution factor $D_{K^0_S \rho \gamma }$ depends on the amplitudes of the two-body decays $\rho(770)^0 K^0_S, ~ K^*(892)^+ \pi^-$ and $  (K\pi)^+_0 \pi^-$~\cite{hebinger}:
\begin{equation}
 D_{K^0_S \rho \gamma} =\frac{\int \bigl [|A_{\rho K^0_S} |^2 -|A_{K^{*+} \pi^-} |^2-|A_{(K \pi)^+_0 \pi^-} |^2 + 2 {\cal R}e (A^*_{\rho K^0_S}A_{K^{*+} \pi^-})   + 2 {\cal R}e (A^*_{\rho K^0_S}A_{(K\pi)^+_0 \pi^-)}   \bigr] dm^2}
{\int \bigl [|A_{\rho K^0_S} |^2 +|A_{K^{*+} \pi^-} |^2 + |A_{(K \pi)^+_0 \pi^-} |^2 + 2 {\cal R}e (A^*_{\rho K^0_S}A_{K^{*+} \pi^-})   + 2 {\cal R}e (A^*_{\rho K^0_S}A_{(K\pi)^+_0 \pi^-)}\bigr] dm^2}
\end{equation}
\noindent
Thus, we need to measure all two-body amplitudes to determine $D_{K^0_S \rho \gamma}$. For this task, we use the $B^+ \ra K^+ \pi^+ \pi^- \gamma$ decay  since due to higher signal yields we achieve more precise measurements and the charged mode is related to
$B^0 \ra K^0_S \pi^+ \pi^- \gamma$ by isospin. The $K \pi \pi $ final state is produced by several kaonic resonances. We extract these from a fit to the $K^+ \pi^+ \pi^- $ mass spectrum.

\section{Determination of the dilution factor $D_{K^0_S \rho \gamma}$ using the decay $B^+ \ra K^+ \pi^+ \pi^- \gamma$ }

We combine a high-energy photon ($1.5 < E_\gamma < 3.5$ GeV) with two charged pions and a charged kaon. The signal event selection is based on two kinematic observables, the beam-energy-constrained mass $m_{ES} =\sqrt{\frac{1}{4} E^2_{CM} -p^{*2}_B}$ and the energy difference $\Delta E=E^*_B-\frac{1}{2}E_{CM}$ where $E_{CM}$ is the center-of-mass energy and $p^*_B, E^*_B$ are momentum and energy of the $B$ meson in the $B$ rest frame. In addition, six event shape observables that are combined into a Fisher discriminant \fisher~\cite{fisher} to separate signal from $e^+ e^- \ra q \bar q$ continuum background ($q =u, d, s, c$). We further use a likelihood ratio to discriminate against photons from $\pi^0$ and $\eta$ decays. Requiring ${\cal L}_{\pi^0} ~({\cal L}_\eta) < 0.86 ~(0.957)$ retains $93\%~(95\%)$ signal and removes $83\%~(87\%) ~ q\bar q$ and $63\%~(10\%)~ B \bar B$ backgrounds. 

\begin{figure}[h]
\centering
\includegraphics[width=74mm]{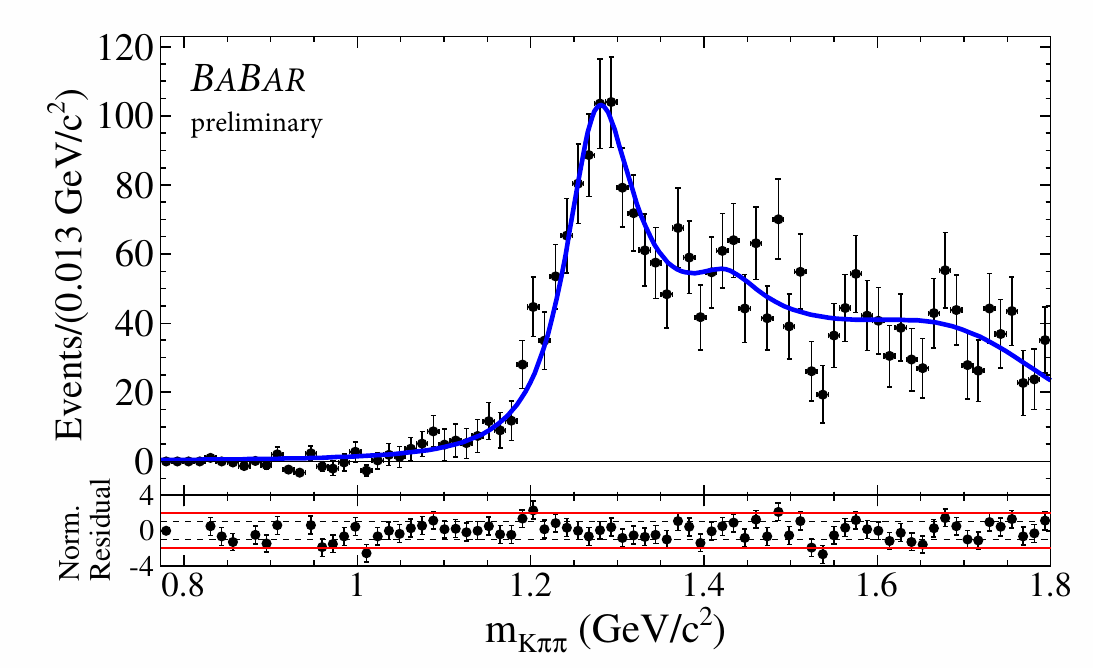}
\includegraphics[width=76mm]{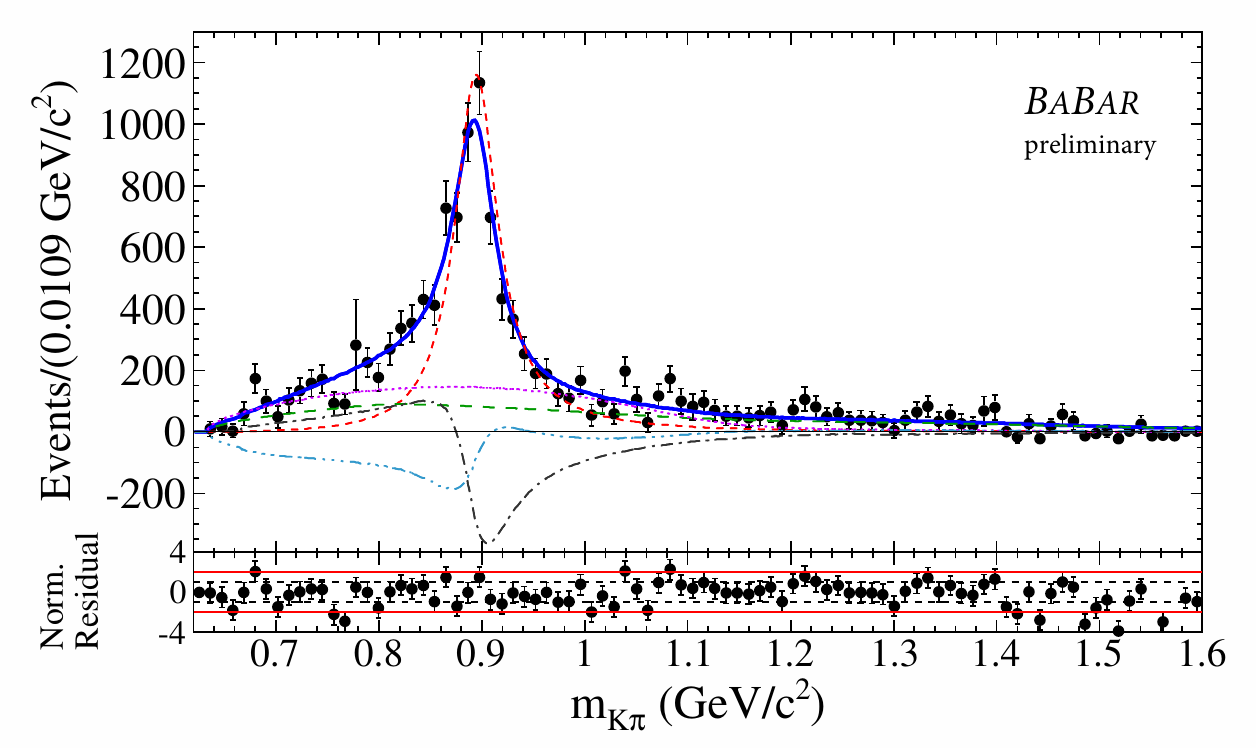}
\caption{The  $m_{K \pi \pi}$ (left) and $m_{K \pi}$ (right) spectra for correctly-reconstructed $B^+ \ra K^+ \pi^+ \pi^- \gamma$ signal events extracted from the maximum likelihood fit to \mes, \DE\ and ${\cal F}$ using the \splot\ technique. Points with error bars represent the sum of sWeights. Blue solid curves show the fit to the $m_{K \pi \pi}$ spectrum and the total PDF fit projection to $m_{K \pi}$, respectively. The red dashed, green long-dashed and magenta dotted lines represent the $K^*(892)$, $\rho(770)^0$ and $(K \pi)_0$ contributions, respectively. The gray dash-dotted and the light blue dash-triple-dotted lines show the interference between $K^*(892)^0$ and $\rho(770)^0$  and that between $(K \pi)_0$ and $\rho(770)^0$, respectively. Note that negative entries are a feature of the \splot\ technique due to background subtraction. The two lower plots show residuals in units of standard deviations. The black dotted and red solid lines indicate the one and two standard
deviation levels, respectively. }
 \label{fig:kpp}
\end{figure}

Using an extended unbinned maximum likelihood fit to \mes, \DE\ and \fisher, we extract a \Bkpp\ signal yield of $2441 \pm 91^{+41}_{-54}$ events for $m_{K \pi \pi} < 1.8~\rm GeV/c^2$. Using the \splot\ technique~\cite{splot}, we extract the \mkpp, \mkp\ and \mpp invariant-mass spectra. 
We model the \mkpp\ invariant-mass spectrum with  five resonances using coherent sums of  same-spin terms ( \Ktwelve +\Kfourteenz, \Kfourteent +\Ksixteen, \Kfourteenth), each parameterized by a relativistic Breit-Wigner line shape. The fit has eight free parameters, the magnitudes of the \Kfourteenz, \Kfourteent, \Ksixteen\ and \Kfourteenth, two relative phases and the widths of the \Ktwelve\ and \Ksixteen. We determine the resonance fit fractions and the interference fit fractions from an 80-bin maximum likelihood fit to the \mkpp\ spectrum. Table~\ref{tab:kresonance} lists the branching fractions of the individual kaonic resonances. Figure~\ref{fig:kpp} (left) shows the \mkpp~ \splot\ spectrum with the fit overlaid. We measure the $B^+ \ra K^+ \pi^+ \pi^- \gamma$ branching fraction to be ${\cal B}(B^+ \ra K^+ \pi^+ \pi^- \gamma)=(27.2\pm 1.0 \pm 1.2)\times 10^{-6}$.

\begin{table}
\centering
\caption{Fit results for amplitudes and phases for different $K^+ \pi^+ \pi^-$ resonances obtained from the maximum likelihood fit to the $m_{K \pi \pi}$ spectrum and computed branching fractions after efficiency correction and division by secondary branching fractions~\cite{pdg}. The first uncertainty is statistical, the second is systematic, and the third, when present, comes from the error of secondary branching
fraction. 
 } 
\vskip 0.3cm
{\footnotesize
\begin{tabular}{|l|c|c|c|c|}
 \hline \hline 
 Mode & Amplitude ~~[Phase (rad)] &${\cal B} (B^+ \ra mode) $ & ${\cal B} (B^+ \ra mode) $ & PDG  \\ 
       &    & ${\cal B} (K_{res} \ra K \pi \pi) $ &$ [ 10^{-6}]$&  $[10^{-6}]$~\cite{pdg}   \\   
       &    & $  [10^{-6}] $  &    & \\  \hline
 
$ B^+ \ra K^+ \pi^+ \pi^- \gamma$ & & & $27.2\pm 1.0 \pm 1.2$ & $27.6 \pm 2.2$ \T \B \\ \hline

$B^+ \ra K_1 (1270)^+ \gamma $ &1.0 (fixed) ~~ [0.0 (fixed)] &$14.5^{+2.0+1.2}_{-1.3-1.2} $& $ 44.0^{+6.0+3.5+4.6}_{-4.0-3.6-4.6} $ & $43\pm 13$ \T \B \\
$B^+ \ra K_1 (1400)^+ \gamma $ &$0.72^{+0.10+0.12}_{-1.0-0.08}~~[2.97^{+0.17+0.11}_{-0.17-0.12}]$ &$~ 4.1 ^{+1.9+1.3}_{-1.2-0.8} $ & $ ~9.6^{+4.6+3.0+0.6}_{-2.9-1.8-0.6} $ & $ < 15~ @ 90\% ~CL $ \T \B \\
$B^+ \ra K^* (1410)^+ \gamma $ &$1.31^{+0.16+0.21}_{-0.16-0.15} ~~[3.15^{+0.12+0.03}_{-0.12-0.04}]$&$ 10.5 ^{+2.1+2.1}_{-1.9-0.9} $ & $ 25.8^{+5.2+5.1+2.6}_{-4.6-2.2-2.6} $ &  $\relbar$ \T \B \\
$B^+ \ra K^*_2 (1420)^+ \gamma $ &$2.07^{+0.28+0.31}_{-0.28-0.23}$ ~~[0.0 (fixed)]& $ ~1.2 ^{+1.2+0.9}_{-1.0-1.2} $ & $ ~8.7^{+8.7+6.2+0.4}_{-7.0-8.5-0.4} $ & $ 14 \pm 4$ \T \B \\
$B^+ \ra K^* (1680)^+ \gamma $ &$0.29^{+0.09+0.08}_{-0.09-0.15}$~~[0.0 (fixed)] &$ 16.6 ^{+1.7+3.6}_{-1.4-2.7} $ & $ 70.0^{+7.2+15+5.7}_{-5.7-11-5.7} $ & $ < 1900 ~@ 90\% ~CL$ \T \B \\
 
\hline \hline                        
\end{tabular}
}
\label{tab:kresonance}
\end{table}

We further perform a 90-bin maximum likelihood fit to the \mkp\ spectrum after correcting for the weighted efficiency in each bin. In the fit we include the $K^*(890)^0$ and a $(K^+ \pi^-)_0$ non-resonant S-wave contribution. Furthermore, we include a contribution from the $\rho^0$ that yields a broad structure in the $K^+ \pi^-$ mass spectrum. We model the $K^*(890)^0$  with a relativistic Breit-Wigner line shape, the  $\rho(770)^0$ with a Gounaris-Sakurai line shape~\cite{GS} and the $(K^+ \pi^-)_0$ with the LASS parameterization~\cite{LASS}. For each resonance, we account for line shape distortions above the pole mass caused by low-mass $K \pi \pi$ resonances. We include interference between  $K \pi$ and $\pi \pi$ P-wave as well as interference  between $K \pi$ S-wave and $\pi \pi$ P-wave. Figure~\ref{fig:kpp} (right) shows the \mkp\ \splot\ spectrum with the fit components superimposed. Table~\ref{tab:kp} lists the branching fraction of the different resonances decaying to $K^+ \pi^-$ and $\pi^+ \pi^-$. This is the first observation of the decay $B^+ \ra K^+ \rho^0 \gamma$ and the $B^+ \ra (K \pi)^{*0}_0 \pi^+ \gamma$ S-wave contribution. 

From the $(K^+ \pi^-)_0$ S-wave component, we separate the $K^* _0(1430)$ resonant contribution by performing separate integration of the resonant part, the non-resonant part and the coherent sum. We find 95.6\% non-resonant, 7.92\% resonant and -3.51\% destructive interference contributions. For the resonant contribution, we extract a branching fraction of $(1.44 \pm 0.19^{+0.26}_{-0.34}\pm 0.14) \times 10^{-6}$ where the first uncertainty is statistical, the second is systematic and the third results from secondary branching fractions. Since in the present analysis the $ K^*_0(1430)$ contribution is modeled to come exclusively from $B \ra K_1(1270) \gamma  \ra K^*_0(1430) \pi \gamma$, we determine the branching fraction to be ${\cal B} (K_1(1270) \ra K^*(1430) \pi) =3.34^{+0.62+0.64}_{-0.54-0.82}\%$. This value is in good agreement with the Belle measurement~\cite{belle11}. Inserting the measured two-body amplitudes into Eqn~(1), we obtain a dilution factor of $D_{K^0_S \rho \gamma} =- 0.79^{+0.18}_{-0.17}$ using the same mass constraints as in the $B^0 \ra K^0_S \pi^+ \pi^- \gamma$ analysis, $m_{K \pi \pi} < 1.8~\rm GeV/c^2$, $0.6 < m_{\pi \pi} < 0.9~\rm GeV/c^2$, $m_{K \pi}< 0.845~\rm GeV/c^2$ and  $m_{K \pi}>0.945~\rm GeV/c^2$.

\begin{table}
\centering
\caption{ Fit results for amplitudes and phases and computed branching fractions for different resonances ($R$) decaying to $h\pi$ ($h=K, \pi$) obtained from the maximum likelihood fit to the $m_{K \pi}$ spectrum after efficiency correction and division by secondary branching fractions~\cite{pdg}.  The first uncertainty is statistical, the second is systematic, and the third (when present) comes from the error of the secondary branching fraction. The last two rows show  resonant and non-resonant parts of the LASS parametrization. 
 } 
\vskip 0.3 cm
{\footnotesize
\begin{tabular}{|l|c|c|c|}
 \hline \hline 
 Mode & Amplitude~~ [Phase (rad)] & ${\cal B} (B^+ \ra Mode)$ $ [ 10^{-6}]$ & PDG  $[10^{-6}]$~\cite{pdg}  \\  \hline
 
$ B^+ \ra K^+ \pi^+ \pi^- \gamma$ &  & $27.2\pm 1.0 \pm 1.2~~~~~~~~~~~~~~~~~~$ & $27.6 \pm 2.2$~~~~~~~~~~~~\T \B  \\ \hline
$K^*(892)^0 \pi^+ \gamma $ & 1.0 (fixed)~~~~~~ [0.0 (fixed)] & $ 26.0^{+1.4}_{-1.3}\pm 1.8 ~~~~~~~~~~~~~~~~~~~~$ & $20^{+7}_{-6}~~~~~~~~~~~~~~~~~~~$ \T \B \\
$K^+ \rho(770)^0 \gamma $ &$0.72^{+0.02+0.01}_{-0.02-0.02}~~ [3.11^{+0.04+0.06}_{-0.04-0.05}]$& $9.2^{+0.8}_{-0.7}\pm 1.3\pm 0.02 $~~~~~~~& $< 20 ~@ 90\%~ CL$ ~~~~~~~~~~\T \B  \\
$(K \pi)^0_0 \pi^+ \gamma $ &$0.82^{+0.04+0.02}_{-0.05-0.05}~~[3.19^{+0.13+0.11}_{-0.13-0.09}]$  & $ 11.3 \pm 1.5^{+2.0}_{-2.6} $  ~~~~~~~~~~~~~~~~~~~&  $\relbar$  \T \B \\ \hline
$(K \pi)^0_0 \pi^+ \gamma $ (NR) &  & $ 10.8^{+1.4+1.9}_{-1.5-2.5}$ ~~~~~~~~~~~~~~~~~~~~~~~& $ < ~9.2 ~@90\% ~CL$ ~~~~~~~~~\T \B  \\
$K_0^* (1680)^0 \pi^+ \gamma $ &  & $ 1.44\pm 0.19^{+0.26}_{-0.34}\pm 0.14 $ &  $\relbar$ \T \B  \\
 
\hline \hline                        
\end{tabular}
}
\label{tab:kp}
\end{table}

\section{Time-dependent analysis of $B^0 \ra K^0_S \pi^+ \pi^- \gamma$}

\begin{figure}[h]
\centering
\includegraphics[width=74mm]{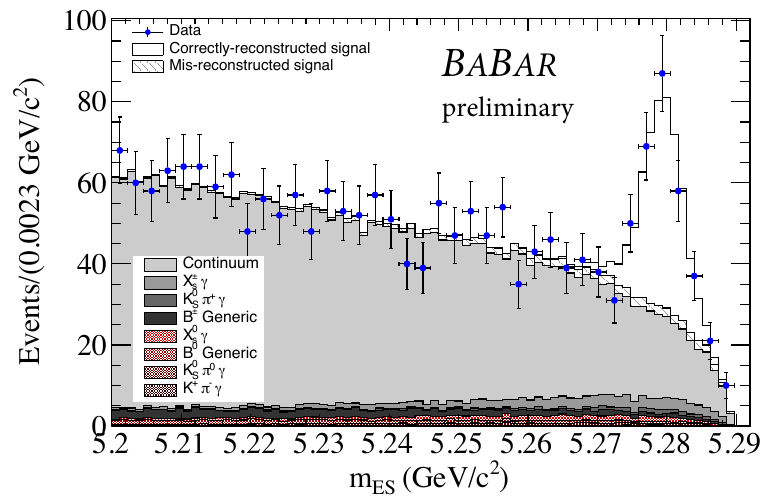}
\includegraphics[width=76mm]{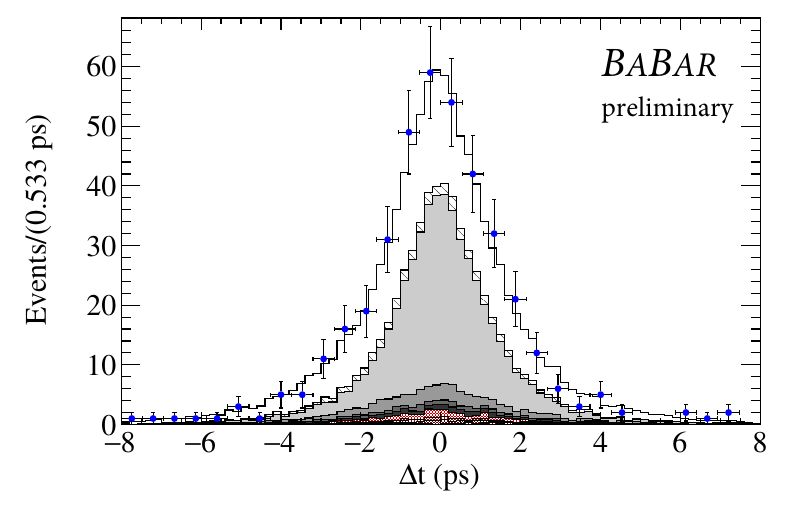}
\caption{The \mes\ (left) and $\Delta  t$ (right) projections of the fit to the $B^0 \ra K^0_S \pi^+ \pi^- \gamma$ sample after requiring $-0.15 < \Delta E < 0.10 ~\rm GeV$ for the $m_{ES}$ and $ \Delta t$ spectra plus $m_{ES}> 5.27~\rm GeV/c^2$ for the $\Delta t$ spectrum. Points with error bars show the data and stacked histograms show the fit projections. Different shadings represent different backgrounds as described in the legend. }
 \label{fig:mes}
\end{figure}

The selection of the $K^0_S \pi^+ \pi^- \gamma$ mode is the same as that for the $K^+ \pi^+ \pi^- \gamma$ mode, except for the replacement of the $K^+$ with a $K^0_S$ requiring $|m_{\pi\pi} -m_{K^0_S}| < 11~ \rm MeV/c^2$, a lifetime significance of more than five standard deviations and $\vec p_{K^0_S} \cdot \vec l_{K^0_S}^{\rm flight}/|\vec p_{K^0_S} \cdot \vec l_{K^0_S}^{\rm flight}| >0.995$ where $\vec p_{K^0_S} $ is the $K^0_S$ momentum and $\vec l_{K^0_S}^{\rm flight}$ is the vector connecting the $B$ and $K^0_S$ decay vertices. We apply the same \mkpp, \mkp\ and \mpp\ mass selections as for the charged mode. For most backgrounds, we use an Argus PDF~\cite{argus} for \mes, Chebychev polymomials for \DE\ and Gaussian or exponential functions for ${\cal F}$. We optimize the selection on ${\cal F}$ to minimize statistical errors of \CP\ parameters. We perform an extended unbinned maximum likelihood fit to extract the \Bkspp\ signal yield along with time-dependent \CP\ asymmetry parameters $S_{K^0_S \pi \pi \gamma}$ and $C_{K^0_S \pi \pi \gamma}$. The likelihood function for event $i$ is a sum over class $j$ (signal and backgrounds) in which each PDF depends on \mes, \DE, ${\cal F}$ and $\Delta t$. For most classes we can factorize the likelihood in the following way: 

\begin{equation}
{\cal P}^i_j (m_{ES}, \Delta E, {\cal F}, \Delta t, \sigma_{\Delta t}, q_{tag}, c) = {\cal P}^i_j (m_{ES}) {\cal P}^i_j (\Delta E) {\cal P}^i_j ({\cal F}) {\cal P}^i_j (\Delta t, \sigma_{\Delta t}; q_{tag},c)
\end{equation}
\noindent
where $q_{tag} = +1 (-1) $ for $B_{tag} =B^0 (\bar B^0)$ and $c$ represents the tagging category. We use six mutually exclusive tagging categories and collect all non-tagged events in a seventh category. We parametrize the proper time distribution for $B^0 \ra K^0_s \rho^0 \gamma$ events for tagging category $c$ by:

\begin{equation}
{\cal P}^i_{sig} (\Delta t, \sigma_{\Delta t}; q_{tag},c) =\frac{\exp [-\frac{|\Delta t |}{\tau_{B^0}}]}{4\tau_{B^0}}\bigl[ 1+ \frac{1}{2}q_{tag}\Delta D_c +q_{tag} <D>_c\bigl ( S \sin(\Delta m_{B^0_d} \Delta t) - C \cos(\Delta m_{B^0_d} \Delta t) \bigr) \bigr] $$
$$\otimes ~R^c_{sig} (\Delta t, \sigma_{\Delta t}) ~~~~~~~~~~~~~~~~~~~~~~~~~~~~~~~~~~~~~~~~~~~~~~~~~~~~~~~~~~~~~~~~~~~
\end{equation}
\noindent
where $<D>_c$ is the average $B^0 \bar B^0$ tagging dilution for category $c$ and $\Delta D_c$ is the difference in $D_c$ between $B^0$ and $\bar B^0$ tags~\cite{babar05, babar07}. We use the tagging algorithm and $<D>_c$, $\Delta D_c$ values for the six tagging categories from the $B^0 \ra (c \bar c) K^{(*)0}$ analysis~\cite{babar07}. 
The seventh category of untagged events is useful for the determination of the direct \CP\ asymmetry~\cite{gardner}. We add $\Delta t$ background PDFs for charged $B$ decays, $B^0$ decays to flavor eigenstates, $B^0$ decays to \CP\ eigenstates as well as those for $q \bar q$ contributions. Figure~\ref{fig:mes} (left) shows the \mes\ projection of the fit. For $|\Delta t| > 20~\rm ps$ and $\sigma_{\Delta t} < 2.5~\rm ps$, the maximum likelihood fit yields $246\pm 24 ^{+14}_{16}$ signal events from which we compute a branching fraction of ${\cal B}(B^0 \ra K^0_S \pi^+ \pi^- \gamma = (24.0 \pm 2.4 ^{+1.7}_{-1.8}) \times 10^{-6}$. 

Figure~\ref{fig:mes} (right) shows the $\Delta t$ distribution, which looks rather symmetric indicating small \CP\ violation as expected in the SM. 
The time-dependent \CP\ parameters are measured to be
$ S_{K^0_S \pi^+ \pi^- \gamma}=0.14 \pm 0.25^{+0.04}_{-0.03}$ and $ C_{K^0_S \pi^+ \pi^- \gamma}=-0.39 \pm 0.20\pm 0.05$.
After correcting for $ D_{K^0_S \rho \gamma}$, we measure a mixing-induced \CP\ parameter of $S_{K^0_S \rho^0 \gamma}=-0.17 \pm 0.32^{+0.07}_{-0.06}$. Our result is consistent with the Belle result~\cite{belle08}. Both results are consistent with zero and thus agree with the SM prediction.

\section{Conclusion}

We observed the decays $B^+ \ra K^+ \pi^+\pi^-\gamma$ and $B^0 \ra K^0_S \pi^+\pi^-\gamma$ and measured their branching fractions. In the charged mode, we observed five kaonic resonances decaying to $K \pi \pi $ for which we measured fit fractions and branching fractions. We found first evidence for $B \ra K_1(1400)^+ \gamma$, $B \ra K^*(1410)^+ \gamma$ and  $B \ra K^*(1400)^+ \gamma$. We further determined the dilution factor $D_{K \rho \gamma}$ after measuring amplitudes and phases of $B \ra K^*(892)^0 \pi^+ \gamma$, $B^+ \ra K^+ \rho(770) \gamma$ and $B \ra (K^+ \pi^-)_0 \pi^+\gamma$ modes. The latter two results are first measurements. In the neutral mode, we measured the time-dependent \CP\ asymmetry parameters $S_{K^0_S \pi^+ \pi^- \gamma}$ and $C_{K^0_S \pi^+ \pi^- \gamma}$  and in turn the \CP\ asymmetry parameter $S_{K^0_S \rho^0 \gamma}$ for the $K^0_S \rho(770)^0 \gamma $ \CP\ eigenstate. Presently, experimental uncertainties are too large to set meaningful limits on right-handed photons. This remains a task for Belle II.


\begin{thebibliography}{99}
\bibitem{Atwood}W. Atwood, M. Gronau and A. Soni, Phys. Rev. Lett. {\bf 79}, 185 (1997).
\bibitem{Fujikawa}K. Fujikawa and A. Yamada, Phys. Rev. {\bf D49}, 5890 (1994). 
\bibitem{Babu}K.S. Babu, K. Fujikawa and A. Yamada, Phys. Lett. {\bf B333}, 196 (1994). 
\bibitem{Cho}P. L. Cho and M. Misiak, Phys. Rev. {\bf D49}, 5894 (1994).
\bibitem{babar12}J. P. Lees {\it et al.} (\babar), Phys.Rev.{\bf D86}, 112008 (2012); ibid, Phys.Rev.Lett. {\bf 109}, 191801 (2012).
\bibitem{babar01}B. Aubert (\babar), Phys. Rev. Lett. {\bf 88}, 101805.
\bibitem{belle09}A. Limosani {\it et al.} (Belle), Phys. Rev. Lett. {\bf 103}, 241801 (2009); T. Saito {\it et al.} (Belle), Phys.Rev. {\bf D91}, 5, 052004 (2014). 
\bibitem{belle04}N. Nakao {\it et al.} (Belle), Phys.Rev. {\bf D69}, 112001 (2004).
\bibitem{pdg}K. Olive {\it et al.} (Particle Data Group), Chin. Phys. {\bf C38}, 090001 (2014).
\bibitem{misiak}M. Misiak {\it et al.}, Phys. Rev. Lett. {\bf 98}, 022002 (2007).
\bibitem{hebinger}J. Hebinger {\it et al.}, LAL-15-75; http://publication.lal.in2p3.fr/2015/1550 note-v3.pdf.
\bibitem{fisher}R. A. Fisher, Annals Eugen. {\bf 7}, 179 (1936).
\bibitem{splot}M. Pivk and F.R. Le Diberder, Nucl. Instrum. Meth. {\bf A555}, 356 (2005).
\bibitem{GS}G.J. Gounaris and J.J. Sakurai, Phys. Rev. Lett. {\bf 21}, 244 (1968).
\bibitem{LASS}D. Aston {\it et al.}, Nucl. Phys. {\bf B296}, 493 (1988).
\bibitem{belle11}H. Guler {\it et al.} (Belle), Phys. Rev. {\bf D83}, 032005 (2011).
\bibitem{argus}H. Albrecht (ARGUS), Z. Phys. {\bf C48}, 543 (1990). 
\bibitem{babar05}B. Aubert (\babar), Phys. Rev. Lett.  {\bf 94}, 161803 (2005).
\bibitem{babar07}B. Aubert (\babar), Phys. Rev. Lett.  {\bf 99}, 171803 (2007).
\bibitem{gardner}S. Gardner and J. Tandean, Phys. Rev. {\bf D69}, 034011 (2004). 
\bibitem{belle08}J. Li et al. (Belle), Phys. Rev. Lett. {\bf 101}, 251601 (2008).

\end{thebibliography}
\end{document}